# Mixing properties of room temperature patch-antenna receivers in a mid-infrared (λ ~ 9 μm) heterodyne system

*Azzurra Bigioli\*, Djamal Gacemi, Daniele Palaferri, Yanko Todorov, Angela Vasanelli, Stephan Suffit, Lianhe Li, A. Giles Davies, Edmund H. Linfield, Filippos Kapsalidis, Mattias Beck, Jérôme Faist, Carlo Sirtori\**

\*corresponding author: azzurra.bigioli@lpa.ens.fr, carlo.sirtori@ens.fr

A. Bigioli, D. Gacemi, D. Palaferri, Y. Todorov, A. Vasanelli, C. Sirtori

*Laboratoire de Physique de l'Ecole Normale supérieure, ENS, Université PSL, CNRS, Sorbonne Université, Université de Paris, 24 rue Lhomond, 75005, Paris, France*

S. Suffit

*Laboratoire Matériaux et Phénomènes Quantiques, Université de Paris, CNRS-UMR 7162, 75013 Paris, France*

L. Li, A. G. Davies, E. H. Linfield

*School of Electronic and Electrical Engineering, University of Leeds, LS2 9JT Leeds, UK*

F. Kapsalidis, M. Beck, J. Faist

*ETH Zurich, Institute of Quantum Electronics, Auguste-Piccard-Hof 1, 8093 Zurich, Switzerland*

A room-temperature mid-infrared (λ= 9 μm) heterodyne system based on high-performance unipolar optoelectronic devices is presented. The local oscillator (LO) is a quantum cascade laser, while the receiver is an antenna coupled quantum well infrared photodetector optimized to operate in a microcavity configuration. Measurements of the saturation intensity show that these receivers have a linear response up to very high optical power, an essential feature for heterodyne detection. By an accurate passive stabilization of the local oscillator and minimizing the optical feed-back the system reaches, at room temperature, a record value of noise equivalent power of 30 pW at 9μm.

Finally, it is demonstrated that the injection of microwave signal into our receivers shifts the heterodyne beating over the bandwidth of the devices. This mixing property is a unique valuable function of these devices for signal treatment.

**Introduction**

Mid-infrared (MIR) quantum cascade lasers (QCLs) are today mature, powerful and stable radiation sources that can be used as local oscillators (LO) in heterodyne systems for coherent low-noise detection. [1] QCL based heterodyne systems have been pioneered in the last decade and were mainly intended for environmental and astrophysics applications. [2] As the identification of remote substances requires non-contact instrumentations, a great effort has been devoted to the realization of compact and scalable setups. [3] [4]

LIDAR systems for on-board automotive platforms employ coherent detection in order to monitor distance and velocity of objects approaching vehicles. Furthermore, they can be exploited for studying the velocity of interstellar medium and the atmosphere composition of terrestrial planets. [5] [6] [7] MIR coherent detection paves the way to high-frequency free space communications, without the need to deploy heavy infrastructures. [8] In MIR range and in adverse conditions, such as fog or dusty environment, an electromagnetic signal propagates through the atmosphere much farther than at 1.55 µm securing communications more efficiently. [9] In addition to sensing and communications, few photons coherent non-destructive MIR detection is very promising in quantum metrology, for fundamental researches such as the precise definition of the molecular hyperfine structure and the study of parity violation in molecules. [10] [11] [12]

For all these applications mostly cryogenically cooled mercury-cadmium-telluride (MCT) detectors have been employed for the last 20 years. However, they are limited by material non-uniformity, significant 1/f noise and intrinsic low-speed response. Recently, 2D materials photodetectors have attracted a lot of attention, but they are not yet at the state of the art and moreover far from presenting a mature and reproducible fabrication process. [13] Quantum well infrared photodetectors (QWIPs) based on GaAs/AlGaAs heterostructures have a competitive background-limited detectivity ($10^{10}$ cm $Hz^{1/2}$ $W^{-1}$) and a very high frequency response up to 100 GHz, due to their intrinsic short carrier life-time (few ps). [14] [15] They are therefore ideal candidates for a 9µm heterodyne detection technology with tens of GHz bandwidth. However, these detectors suffer of relatively high dark current that up to now has imposed cryogenic operational temperatures (70 K for the $\lambda \sim 9$ µm). [16] We have recently demonstrated a quantum well infrared photodetector operating at 9µm embedded into a metamaterial made of sub-wavelength metallic patch-antenna resonators, with strongly enhanced performances up to room temperature. [17] By mixing the frequencies of two QCLs, we have measured a high-frequency signal at 1.7 GHz demonstrating that these detectors operate also as heterodyne receivers.

In this letter we present recent progresses on heterodyne detection using room temperature metamaterial QWIP receivers. We measure a noise equivalent power (NEP) of tens of pW, which is an improvement of 5 orders of magnitude with respect to our previous results. [17] This improvement is due to three factors: an optimized active region design; the passive stabilization of our distributed feedback (DFB) QCLs and an improved optical alignment of the local oscillator. Moreover, we show that QWIP receivers, with an applied microwave bias, can also act as frequency mixers. This enables to shift the heterodyne signal at will over the whole device frequency band, unveiling another intrinsic high frequency function that can be obtained with these detectors.

**The device**

The device used in this study is a GaAs/AlGaAs QWIP containing $N_{qw}$ = 8 quantum wells absorbing at a wavelength of 8.9 µm at room temperature (139 meV). [14] It has been designed with a bound-to-bound structure and has been processed into an array of double-metal patch resonators. These resonators provide sub-wavelength electric field confinement and also act as antennas. [17] [18] An exemplary device is shown in figure 1a, where the 50 x 50 µm$^2$ array is the collection area of the detector. The metamaterial permits to separate the photon collection area from the electrical area, which is only 380 µm$^2$. The resonance wavelength is defined by the lateral patch size $s$ according to $\lambda = 2\,s\,n_{eff}$, where $n_{eff}$ = 3.3 is the effective index. [19] The resonance with the peak responsivity of the detector occurs for structures with $s$ = 1.35 µm.

The active region has been optimized with the intention to improve the performances of this type of metamaterial high temperature detectors. The doping density per well has been decreased with respect to our previous work [17] from $n_{2D}$ = 7.0 × 10$^{11}$ cm$^{-2}$ to $n_{2D}$ = 4.0 × 10$^{11}$ cm$^{-2}$, in order to reduce the Fermi energy of 10 meV and consequently increase the activation energy. The associated lower absorption has been compensated employing 8 quantum wells instead of 5. The thickness of the absorbing region, without taking into account the contact layers, is therefore increased from $L$ = 236 nm to $L$ = 339 nm. The thicker active region also reduces the ohmic losses of the cavity resulting in a higher cavity quality factor, $Q$. [20] [19] We recall that the fraction of photons absorbed in the quantum wells is the branching ratio between the intersubband absorption factor $B_{isb}$ and the metal losses, proportional to $1/Q$. A list with the principal parameters is reported on Table 1. It can be noticed that a compromise has been reached between $B_{isb}$ and the metal losses, $1/Q$ in order to avoid either strong coupling ($B_{isb} \rightarrow \infty$) or to limit the acceptance bandwidth of the detector ($Q \rightarrow \infty$). [19]

| Coefficients | Structure in reference [17] | Current active region |
|---|---|---|
| Plasma energy $E_p$ | 47.2 meV | 35.7 meV |
| Intersubband absorption factor $B_{isb}$ | 0.069 | 0.052 |
| Losses quality factor $Q_{loss}$ | 3.4 | 4.7 |

**Table 1** Comparison of plasma energy, intersubband absorption coefficient and losses quality factor for the structure in reference *[17]* and the optimized active region.

## Results

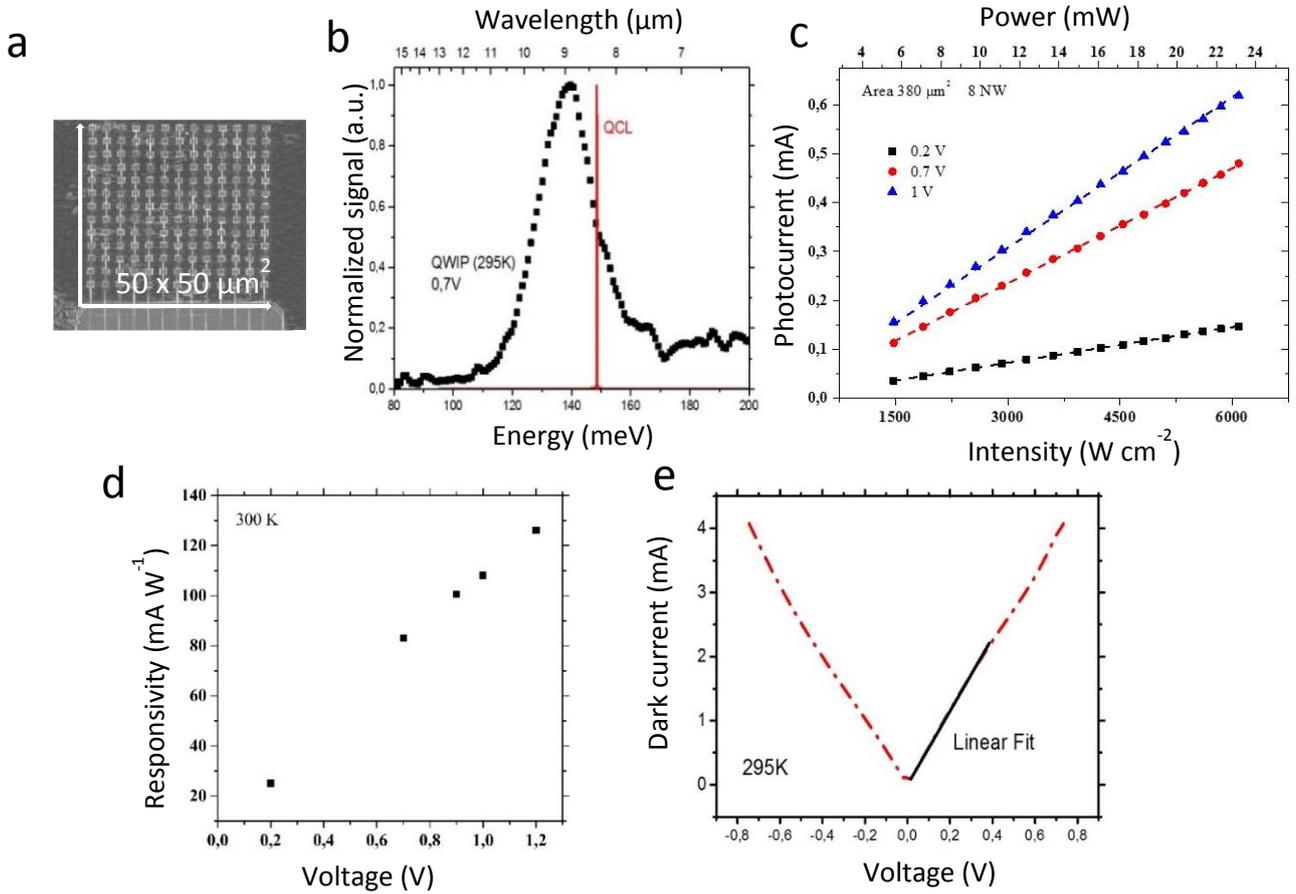

**Figure 1** a) Scanning electron microscope image of the array device. b) Photocurrent spectrum at room temperature with an applied bias of 0.7 V. In red the DFB QCL emission line. c) Photocurrent curves as a function of the incident intensity for three different bias voltages applied to the device. The dashed lines are the linear fit. The density is calculated by dividing the optical power by the electrical area of the device (380 μm$^2$). d) Responsivity of the detector as a function of the voltage applied to the device. e) Dark current curves as a function of voltage at 295 K. The black curve is the linear fit for small voltages.

The active region optimization has led to a room temperature peak detectivity higher of a factor 1.5 ($D^* = 3.3 \times 10^7$ cm Hz$^{1/2}$ W$^{-1}$), with respect to the previous reported value ($D^* = 2.2 \times 10^7$ cm Hz$^{1/2}$

W$^{-1}$). [17] The measurement has been performed using a dedicated setup with a calibrated blackbody source at 1273 K. A room temperature normalized spectrum at 0.7V is presented in Figure 1b.

The benefit of working at room temperature relies not only on important technical advantages, as the disposal of cryogenic cooling systems, but also on a different detector operation condition that changes the physics of the device and very importantly enables extremely high saturation intensities. In Figure 1c we report responsivity curves as a function of the incident power supplied by a QCL, for three different biases applied to the device. No saturation effects appear even at intensities of several kW cm$^{-2}$, which exceed greatly typical saturation intensity of QWIP structures operating at low temperature. [21] [22] Notably these values are far above the saturation intensity of MCT detectors. As a comparison, we have characterized with the same set-up a Vigo MCT (model PVI-4TE-10.6), showing that it saturates at about 400 µW for an area of 0.5 mm x 0.5 mm, corresponding to 0.16 W cm$^{-2}$ only.

The increased saturation intensity at room temperature can be explained in terms of electric field homogeneity, following the model described by Ershov et al. in [21]. At high temperature the generation rate of electrons from the quantum wells to the continuum states and the thermionic transport through the contact barrier increase and therefore a drift current regime always supplies the carriers necessary to refill the wells depopulated by photoemission. No charge depletion in the first few wells occurs; hence the electric field distribution is constant and uniform along the 8 wells at room temperature.

In Figure 1d we show the responsivity measured at room temperature as a function of the applied bias. At room temperature and with an applied bias of 0.7 V the detectors show a relatively high peak responsivity of 80 mA W$^{-1}$. In these operating conditions the dark current is $I_{dark}$ = 3.7 mA (Figure 1e) and, in direct detection, the noise equivalent power (NEP) is in the µW range. However, due to their linear behavior up to exceptionally high power and their very wide frequency response, these detectors have their best performance as heterodyne receivers, where the signal to be detected is multiplied by the LO power. Mounted in our present heterodyne setup, the detectors have shown a NEP of 30 pW with 3 mW of LO power, the maximum obtainable in our configuration. In this condition the LO photocurrent is $I_{LO}$ = 0.2 mA, one order of magnitude smaller than $I_{dark}$, which is the dominant noise contribution.

In our heterodyne experiment the optical power from each DFB quantum cascade laser is collected by $f_{0.5}$ lenses. The two beams are then made collinear using a 50% beam-splitter (BS) and finally focused onto the patch antenna QWIP detector with a $f$ = 6 mm lens. The QWIP is biased with a Keithley 2450 source meter, while the heterodyne signal from the detector is sent through a bias tee

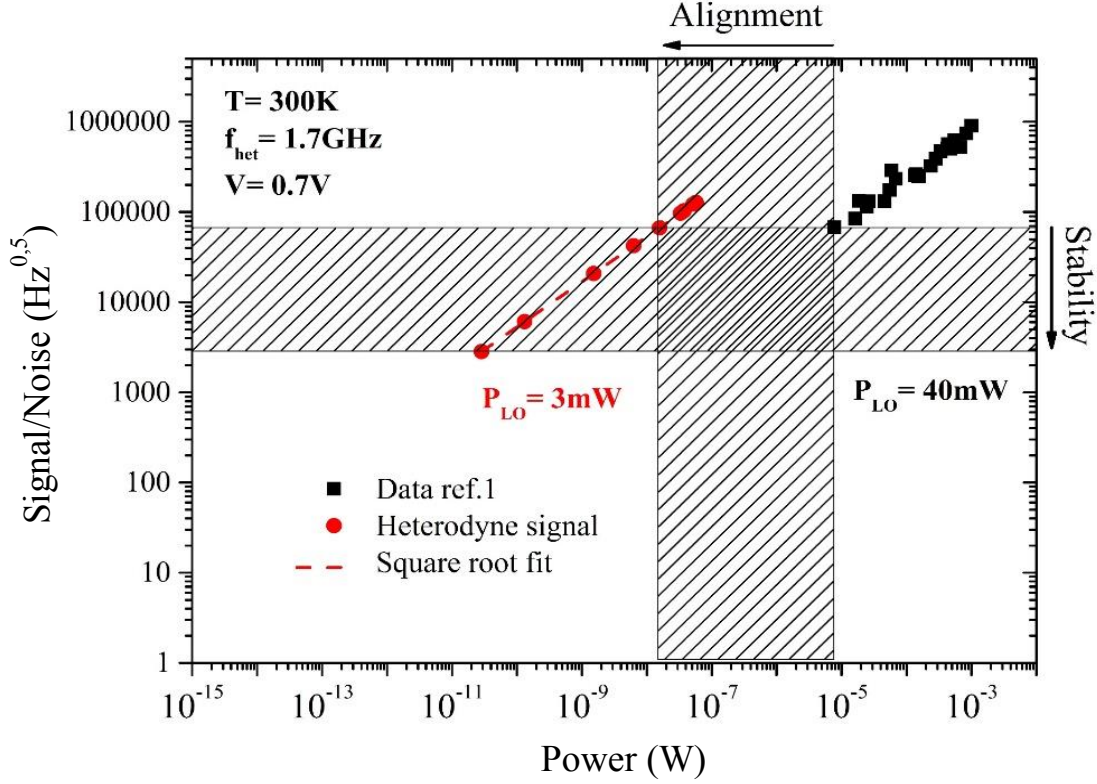

**Fig. 2** log-log plot of the signal-to-noise ratio as function of the QCL power. The noise of the QWIP is calculated using the measured gain and dark current values at room temperature. Black points are data from *[17]*, while red points are the current measurements. The improvement was possible in the horizontal scale thanks to a greater responsivity, in the vertical scale to a better stability.

to the spectrum analyzer Agilent E4407B. The QWIP is mounted on a copper holder without any cooling system. The temperature and current of the lasers are chosen to set the heterodyne signal at a frequency close to 1.7 GHz, well below the detector frequency cutoff of 5 GHz. The bandwidth is at present limited by an impedance mismatch between the detector and the external circuit.

Figure 2 shows a comparison between measurements in the current setup, red points, and the data from our previous work, black symbols. [17] In this graph the *y*-axis is the signal to noise ratio (*S/N*) multiplied by the square root of the integration bandwidth $\sqrt{\Delta f}$, while on the *x*-axis one can read the incident power from the signal-laser, $P_S$. The noise of the QWIP, in a unit frequency band, is calculated using the measured gain and dark current at room temperature. The lowest detectable power on the x-axis is the NEP of the system and, for each set of data, corresponds to a *S/N* = 1 at a given $\sqrt{\Delta f}$ readable on the y-axis. The measured NEP is 30 pW ($\Delta f$ = 9 MHz) for the present setup and was 8 μW ($\Delta f$ = 4.5 GHz) previously, corresponding to an increase in sensitivity of 5 orders of magnitude. [17] Notice that the measurements in reference [17] were taken with a local oscillator power $P_{LO}$ = 40 mW, while in the present configuration the LO power is at most $P_{LO}$ = 3mW, due to optical elements added on the beam path.

The reduction of the integration bandwidth $\Delta f$, thus the stability of the system, has been one of the main issues of our work, that we have addressed by a passive stabilization of the laser frequencies.

To this end we have minimized all the external sources of noise that compromise lasers stability: most importantly i) fluctuations of the lasers temperature, ii) the noise from the lasers current generator and iii) the optical feedback. [23] Temperature was controlled by mounting the DFB lasers on Peltier cooler elements stabilized to the mK, while current noise was minimized by using two low noise homemade current drivers (100 pA Hz$^{-1/2}$). [24] Optical feedback was also reduced by employing an Innovation Photonics MIR isolator based on a Faraday rotator principle. All these arrangements enabled an integration time of ~100 ns ($\Delta f$ = 9 MHz) to be compared to ~ 200 ps ($\Delta f$ = 4.5 GHz) of the work in reference [17]. The reduction of the technical noise gave us a stability which is comparable to the product of the linewidths of the two DFB QC lasers, reported to be 1 MHz each in the literature. [23] Finally, we remark that if we could integrate on a 1 Hz bandwidth, we can extrapolate a theoretical NEP at 0.7V of ~$10^{-18}$ W at 300K.

The lower NEP of the improved setup has been obtained not exclusively from an increased stability of the system, but also from a better alignment of the LO which enables a greater responsivity. This can be readily observed in Figure 2a, where the enhanced stability corresponds to a reduced $\Delta f$ (shift in the vertical direction) while a greater responsivity gives a shift on the horizontal axis. We recall that at optical and infrared frequencies, careful alignment between the LO and signal beams is necessary in order to maintain a constant phase over the surface of the photodetector. [25] The overlap of wave fronts between signal and LO beam is regulated by the relation $A_d \Omega = \lambda^2$, where $A_d$ is the detector area and $\Omega$ is the solid angle of incidence of the radiation and $\lambda$ is the wavelength. At $\lambda$ =8.9 µm and for our detector area, $A_d$ =2500 µm$^2$, the photomixing is assured for a maximum angular tolerance of 32 mrad.

**Frequency shifting of the heterodyne signal by mixing**

The improved stability of our system has allowed us to explore other important functions of these ultrafast receivers. In particular, we have investigated the ability of frequency shifting the heterodyne signal, which is of great interest for signal processing and for future active stabilization of the beating.

This function has been obtained by injecting into the detector a microwave excitation through a -10 dB directional coupler so to modulate the heterodyne signal within the receiver itself, with or without

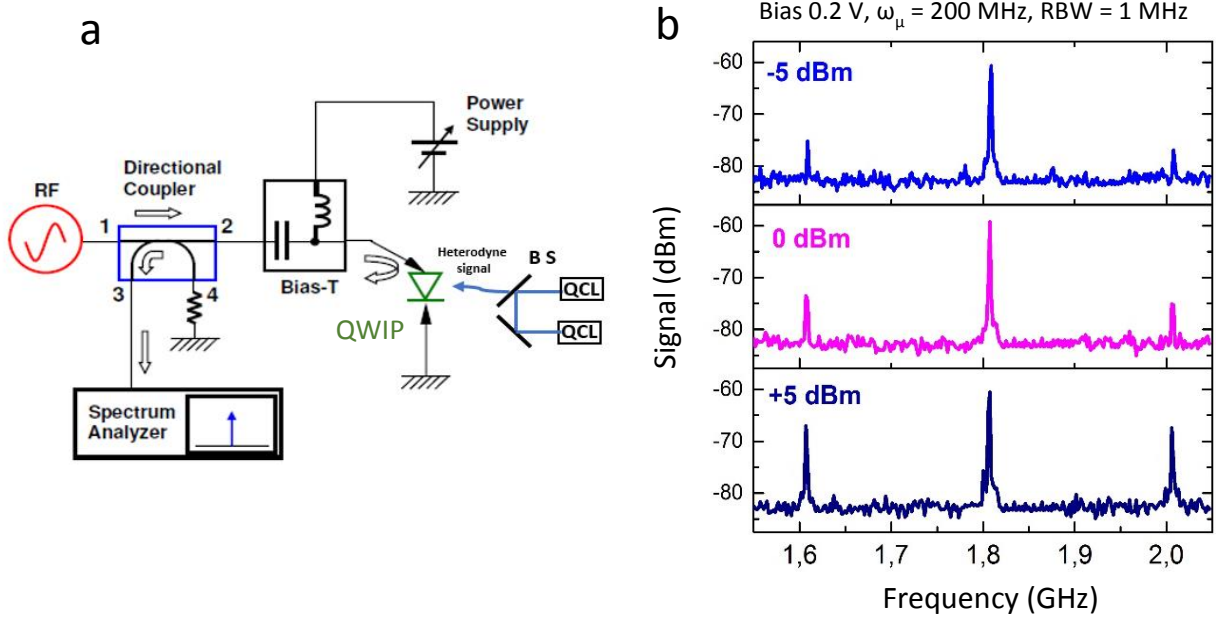

**Figure 3** a) Schematic of the microwave modulation experiment, where the heterodyne signal produced by the two DFB lasers is modulated by a microwave source and biased through a bias T. The signal is recorded by a spectrum analyzer. b) Spectrum analyzer signals with an applied bias of V= 0.2 V, for three different injected powers. The heterodyne signal has a frequency of 1.8 GHz, and two sidebands, generated from the modulation, are distant ±200 MHz from it.

an applied dc bias. A schematic of the set-up is shown in Figure 3a. The QWIP is considered ending in a 50-Ω transmission line and can be modeled by a parallel resistance-capacitance equivalent circuit. [15]

The QWIP current at a fixed voltage $V_0$ is modulated through a small applied microwave voltage $v_\mu = v\,cos(\omega_\mu t)$ around a working point $I_0$. The injected ac bias at frequency $\omega_\mu$ causes also an amplitude modulation of the device responsivity $R(V_0 + v_\mu)$, which allows us to expand the heterodyne current at the first order as:

$$I_{het} = 2R(V_0 + v_\mu)\sqrt{P_{LO}P_s}\,cos(\omega_h t) =$$

$$= I_{het}(\omega_h) + 2v\,\frac{\partial R(V_0)}{\partial v}\sqrt{P_{LO}P_s}\,cos(\omega_h t)\,cos(\omega_\mu t) = I_{het}(\omega_h) + I_{het}(\omega_h \pm \omega_\mu) \quad (1)$$

The heterodyne current is therefore composed of a signal at the frequency $\omega_h$ and two sidebands at $\omega_h \pm \omega_\mu$. The component at $\omega_h$ disappears if no bias is applied to device, as $R(0) = 0$.

Figure 3b shows measurements recorded with a spectrum analyzer of a heterodyne signal at 1.8 GHz. The detector has a constant applied bias $V_0 = 0.2$ V, while the microwave power is increasing top down in the figure. Two sidebands appear at $\omega_h \pm 200$ MHz, corresponding to the frequency of the injected microwave. The signal power of the sidebands is:

$$8R_L^2\left(\frac{\partial R(v)}{\partial V}\right)^2 P_u P_{LO} P_s \quad (2)$$

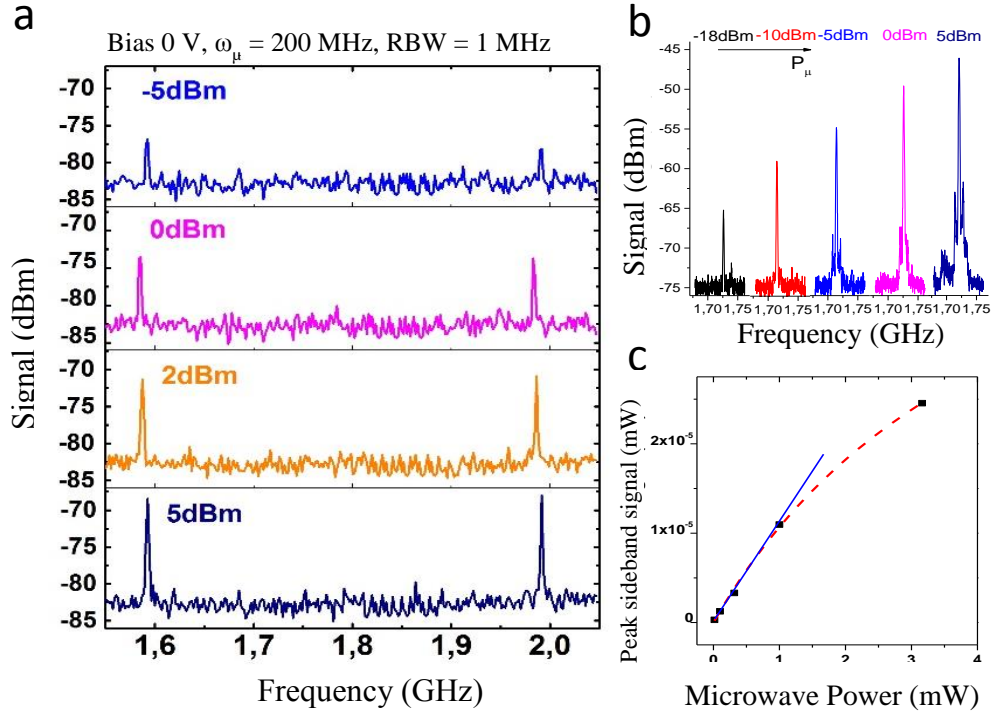

**Figure 4** a) Spectrum analyzer signals with no applied bias, for four different injected powers. The heterodyne signal is no more visible, and two sidebands, generated from the modulation, are present at ±200 MHz from 1.8 GHz. b) Sideband signal at increasing injected power from -18 dBm to 5 dBm. c) Sideband peak signal amplitude as a function of the injected microwave power in linear scale. The blue line is the linear fit, whereas the red dashed line is the fit considering saturation.

where $R_L$ is the load resistance and $P_u$ is the applied microwave power. From this formula, we can infer that an amplification effect of the heterodyne signal to sidebands takes place as a function of the steepness of the responsivity-voltage curve. This is very clear for the unbiased situation, as presented in Figure .

In Figure a we report microwave spectra with no bias applied on the detector, while the injected microwave power increases from -5 dBm to 5 dBm. The heterodyne signal disappears and only the sidebands are observable as predicted by the formula (2). Figure 4b shows the sideband intensity for increasing microwave power on a wide range from Pμ= -18 dBm up to 5 dBm. A linear behavior, as depicted in Figure c, is observed up to 1 mW (blue straight line), which can be converted into a voltage swing of 0.3V on the device, in good agreement with the responsivity curve reported on Figure 1d.

These measurements illustrate that our QWIP heterodyne receiver acts also as a mixer, enabling to shift the heterodyne frequency to sidebands at any frequency within the bandwidth of the detector.

**Conclusions**

We designed an optimized active region to boost the performances of patch-antenna QWIP detectors. The high temperature prevents the formation of space charge domains and a homogeneous electric field along the structure enables a regime of very high saturation intensity. This is essential in heterodyne measurements where powerful LOs are necessary to enter in the low noise regime of heterodyne detection. The use of low noise QCL current drivers, Peltier coolers and limited optical feedback, permits an easier alignment procedure and a longer integration time. With these arrangements the minimum detected power is now tens of pW. Further advances will include the active stabilization of the frequency of the DFB QCLs up to Hz range. This will allow to integrate for longer time scales thus bringing the NEP of the system in the aW regime. [26] Peltier cooling of the detector, high frequency adaptation of its electrical circuit and balanced detection scheme will be also developed to take full benefit of the high frequency and low noise properties of heterodyne technique.

Finally, we demonstrated that this device can be modulated by injected microwave power producing sidebands of a stabilized heterodyne signal that can be swept on a very large frequency band. This could be a practical solution for signal processing in compact QCL-based systems.

**Aknowledgements**

We acknowledge financial support from the Qombs Project (grant agreement number 820419).

**References**

[1] L. Consolino, F. Cappelli, M. Sicialiani de Cumis and P. De Natale, "QCL-based frequency metrology from the mid-infrared to the THz range," *Nanophotonics,* vol. 8, no. 2, pp. 181-204, 2019.

[2] D. Weidemann, F. Tittel, T. Aellen, M. Beck, D. Hosfetter, J. Faist and S. Blaser, "Mid-Infrared trace-gas sensing with a quasi-continuous-wave Peltier-cooled distributed feedback quantum cascade lasers," *Applied Physics B: Lasers and Optics,* vol. 79, no. 7, pp. 907-913, 2004.

[3] T. Strangier, G. Sonnabend and M. Sornig, "Compact setup of a Tunable Heterodyne Spectrometer for Infrared Observations of Atmospheric Trace-Gases," *Remote Sensing,* vol. 5, pp. 3397-3414, 2013.

[4] Z. Su, Z. Han, P. Becla, H. Lin, S. Deckoff-Jones, K. Richardson, L. Kimerling, J. Hu and A. Agarwal, "Monolithic on-chip mid-IR methane gas sensor with waveguide-integrated detector," *Applied Physics Letters,* vol. 114, no. 051103, 2019.

[5] A. Samman, L. Rimai, J. McBride, R. Carter, W. Weber, C. Gmachl, F. Capasso, A. Hutchinson, D. Sivco and A. Cho, "Potential Use of Near, Mid and Far Infrared Laser Diodes in Automotive LIDAR applications," in *IEEE VTS Fall VTC2000. 52nd Vehicular Technology Conference*, 2000.

[6] R. Schieder, D. Wirtz, G. Sonnabend and A. Eckart, "The Potential of IR-Heterodyne Spectroscopy," in *The Power of Optical/IR Interferometry: Recent Scientific Results and 2nd Generation Instrumentation*, 2008.

[7] C. Straubmeier, R. Schieder, G. Sonnabend, D. Wirtz, V. Vetterle, M. Sornig and V. Eckart, "Tunable Heterodyne Receveirs-A promising outlook for Future Mid-Infrared Interferometry," in *Exploring the cosmic frontier*, Springer, 2007.

[8] M. Piccardo, M. Tamagnone, B. Schwarz, P. Chevalier, N. Rubin, Y. Wang, C. Wang, M. Connors, D. McNulty, A. Belyanin and F. Capasso, "Laser radio transmitter," *arXiv:1901.07054 [physics.app-ph]*, 2019.

[9] A. Delga and L. Leviander, "Free-space optical communiations with quantum cascade lasers," in *Quantum Sensing and Nano Electronics and Photonics XVI, SPIE OPTO*, San Francisco, 2019.

[10] S. Borri, G. Insero, G. Santambrogio, D. Mazzotti, F. Cappelli, I. Galli, G. Galzerano, M. Marangoni, P. Laporta, V. Di Sarno, L. Santamaria, P. Maddaloni and P. De Natale, "High-precision molecular spectroscopy in the mid-infrared using quantum cascade lasers," *Applied Physics B,* vol. 125, no. 18, 2019.

[11] S. Cahn, J. Ammon, E. Kirilov, Y. Gurevich, D. Murphee, R. Paolino, D. Rahmlow, M. Kozlov and D. DeMille, "Zeeman-Tuned Rotational Level-Crossing Spectroscopy in a Diatomic Free Radical," *Physical Review Letters,* vol. 112, no. 163002, 2014.

[12] P. Asselin, Y. Berger, T. Huet, L. Margulès, R. Motiyenko, R. Hendrincks, M. Tarbutt, S. Tokunaga and B. Darquié, "Characterising molecules for fundamental physics:an accurate spectroscopic model of methyltrioxorhenium," *Physical Chemistry Chemical Physics,* vol. 19, no. 4576, 2017.

[13] S. Capmakyapan, P. Keng Lu, A. Navabi and M. Jarrahi, "Gold-patched graphene nano-stripes for high-responsivity and ultrafast photodetection from the visible to infrared," *Light: Science & Applications,* vol. 7, no. 20, 2018.

[14] B. Levine, K. Choi, C. Bethea, J. Walker and R. Malik, "New 10 μm infrared detector using intersubband absorption in resonant tunneling GaAlAs superlattices," *Applied Physics Letters,* vol. 50, no. 1092, pp. 1092-1094, 1987.

[15] H. Schneider and H. Liu, Quantum Well Infrared Photodetectors: Physics and Applications, Springer, 2007.

[16] A. Rogalski, "Review Infrared detectors: status and trends," *Progress in Quantum Electronics,* vol. 29, pp. 59-210, 2003.

[17] D. Palaferri, Y. Todorov, A. Bigioli, A. Mottaghizadeh, D. Gacemi, A. Calabrese, A. Vasanelli, L. Li, A. Davies, E. Lienfield, F. Kapsalidis, M. Beck, J. Faist and C. Sirtori, "Room-temperature nine-μm-wavelength photodetectors and GHz-frequency heterodyne receivers," *Nature,* vol. 556, no. 85, 2018.

[18] Y. Todorov, L. Tosetto, J. Teissier, A. Andrews, P. Klang, R. Colombelli, I. Sagnes, G. Strasser and C. Sirtori, "Optical properties of metal-dielectric-metal microcavities in the THz frequency range," *Optics Express,* vol. 18, no. 13, 2010.

[19] D. Palaferri, Y. Todorov, A. Mottaghizadeh, G. Frucci, G. Biasol and C. Sirtori, "Ultra-subwavelength resonators for high temperature high high performance quantul detectors," *New Journal of Physics,* vol. 18, no. 113016, 2016.

[20] C. Feuillet-Palma, Y. Todorov, A. Vasanelli and C. Sirtori, "Strong near field enhancement in THz nano-antenna arrays," *Scientific Reports,* vol. 3, no. 1361, 2013.

[21] E. Ershov, H. Liu, M. Buchanan, Z. Wasilewski and V. Ryzhii, "Photoconductivity nonlinearity at high excitation power in quantum well infrared photodetectors," *Applied Physics Letters,* vol. 70, no. 414, 1997.

[22] C. Mermelstein, H. Schneider, A. Sa'ar, C. Schonbein, M. Walther and G. Bihlmann, "Low-power photocurrent nonlinearity in quantum well infrared detectors," *Applied Physics Letters,* vol. 71, no. 2011, 1997.

[23] S. Schilt, L. Tombez, G. Di Domenico and D. Hofstetter, "Frequency Noise and Linewidth of Mid-infrared Continuous-Wave Quantum-Cascade Lasers: An Overview," in *The Wonders of Nanotechnology: Quantum and Optoelectronic Devices and Applications*, 2013, pp. 261-287.

[24] *The current driver was supplied by the laboratory LPL, Laboratoire de Physique de Lasers Paris 13.*

[25] A. Siegman, "The Antenna Properties of Optical Heterodyne Receveirs," *Applied Optics,* vol. 5, no. 10, 1966.

[26] B. Argence, B. Chanteau, O. Lopez, D. Nicolodi, M. Abgrall, C. Chardonnet, C. Daussy, B. Darquié, Y. Le Coq and A. Amy-Klein, "Quantum cascade laser frequency stabilisation at the sub-Hz level," *Nature Photonics,* vol. 9, pp. 456-460, 2015.